\documentclass[aps,pra,showpacs,twocolumn,superscriptaddress]{revtex4-1}
\usepackage[latin9]{inputenc}
\usepackage{graphicx,color}
\usepackage{verbatim}
\usepackage{units}
\usepackage{bm}
\usepackage{amstext}
\usepackage{amsmath,amssymb}

\newcommand{\tr}{\mathop{\text{tr}}\nolimits}
\definecolor{dgreen}{rgb}{0,0.5,0}
\definecolor{delete}{cmyk}{0.5,0,0,0}

\makeatletter
\@ifundefined{textcolor}{}
{%
 \definecolor{BLACK}{gray}{0}
 \definecolor{WHITE}{gray}{1}
 \definecolor{RED}{rgb}{1,0,0}
 \definecolor{GREEN}{rgb}{0,1,0}
 \definecolor{BLUE}{rgb}{0,0,1}
 \definecolor{CYAN}{cmyk}{1,0,0,0}
 \definecolor{MAGENTA}{cmyk}{0,1,0,0}
 \definecolor{YELLOW}{cmyk}{0,0,1,0}
}

\makeatother

\begin{document}
\title{Identifiability of Open Quantum Systems}
\author{Daniel Burgarth}
\affiliation{Department of Mathematics and Physics, Aberystwyth University, SY23 3BZ Aberystwyth, United Kingdom}
\author{Kazuya Yuasa}
\affiliation{Department of Physics, Waseda University, Tokyo 169-8555, Japan}
\date[]{January 20, 2014}

\begin{abstract}
We provide a general framework for the identification of open quantum systems.
By looking at the input-output behavior, we try to identify the system inside a black box in which some Markovian time-evolution takes place.
Due to the generally irreversible nature of the dynamics, it is difficult to assure full controllability over the system.
Still, we show that the system is identifiable up to similarity under a certain rank condition.
The framework also covers situations relevant to standard quantum process tomography, where we do not have enough control over the system but have a tomographically complete set of initial states and observables.
Remarkably, the similarity cannot in general be reduced to unitarity even for unitary systems, and the spectra of Hamiltonians are not identifiable without additional knowledge.
\end{abstract}

\maketitle

\textit{Introduction.---}
It is well known that quantum process tomography \cite{ref:QuantumEstimation} becomes inefficient
as the dimension of the underlying system increases. In particular,
highly precise controls of the system are required for state preparation
and measurement. It is interesting to consider the case where such
controls are not given, and therefore, in general, the system cannot
be fully characterized. 
In \cite{ref:QSI}, we characterized the unitary quantum
systems which cannot be distinguished by their input-output behavior,
provided that the dynamics of the control system is rich enough to
be ``accessible'' (that is, the system Lie algebra has the maximal rank, and therefore in the unitary case the system is fully controllable).
We found that under such conditions the quantum system can be identified
up to unitary equivalence. This implies that for accessible 
unitary systems the relevant mathematical object which characterizes how
well they can be identified in principle is the Lie algebra $\mathfrak{l}_\text{known}$
generated by a priori information, i.e., known measurements, states, or
Hamiltonians. An accessible unitary system is fully identifiable if
and only if $\mathfrak{l}_\text{known}$ is irreducible. If it is
not irreducible, its commutant characterizes the unidentifiable parameters.

The previous work \cite{ref:QSI} left unanswered such characterization
for open quantum systems, as well as for non-accessible cases, which is relevant for quantum process tomography. In
the present paper we aim at such generalizations. We focus on non-unital
Markovian dynamics and find that under a certain rank condition systems
with equal input-output behavior are related through similarity
transformations. The rank condition describes accessible systems, tomographically complete systems, or all intermediate situations. Remarkably, the similarity cannot in general be reduced to unitarity, even for unitary systems.
We provide examples for such situations and discuss its physical consequences.

\begin{figure}
\includegraphics[width=0.32\textwidth]{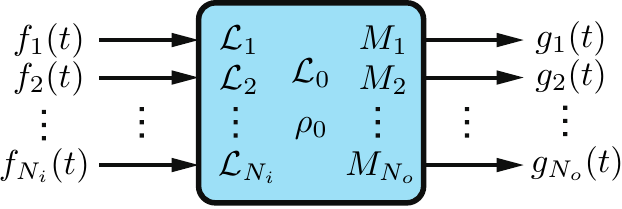}
\caption{Our problem is to identify the system $\sigma=(\mathcal{L}_0,\mathcal{L}_k,\rho_0,M_\ell)$ in the black box, by looking at its response $g_\ell(t)=\tr\{M_\ell\rho(t)\}$ to the inputs $f_k(t)$.}
\label{fig:BlackBox}
\end{figure}
\textit{Setup.---}
We consider a black box with $N_i$ inputs and $N_o$ outputs (Fig.\ \ref{fig:BlackBox}) \cite{ref:QSI,ref:Sontag,ref:GutaYamamoto}.
Inside the black box, some quantum-mechanical dynamics takes place.
Our goal is to find a model for the black box by looking at its input-ouput behavior.
In \cite{ref:QSI}, we studied this problem under the restriction that the dynamics occurring in the black box is known to be unitary.
We here relax this condition and establish a framework that allows us to deal with open quantum systems.

More specifically, we consider a $d$-dimensional system whose dynamics is governed by a master equation
\begin{equation}
\dot{\rho}(t)=\mathcal{L}_{0}(\rho(t))+\sum_{k=1}^{N_{i}}f_{k}(t)\mathcal{L}_{k}(\rho(t)),\quad
\rho(0)=\rho_0,
\label{master}
\end{equation}
that is, the system is assumed to undergo a Markovian time-evolution from some initial state $\rho_0$.
The functions $f_k(t)$ ($k=1,\ldots,N_i$) are our inputs, which are non-negative
\cite{note:1} 
and piecewise constant, while $\mathcal{L}_k$ ($k=0,1,\ldots,N_i$) are some generators of the Lindblad form \cite{ref:DynamicalMap-Alicki}.
The outputs are the expectation values $g_\ell(t)=\tr\{M_\ell\rho(t)\}$ of some  observables $M_\ell$ ($\ell=1,\ldots,N_o$) 
\cite{note:2}. 
We assume that many copies of the system are at hand, so that we can measure the expectation values $g_\ell(t)$ without taking the measurement back-action into account.
Our objective is to identify the system $\sigma=(\mathcal{L}_0,\mathcal{L}_k,\rho_0,M_\ell)$, by looking at the response $\{g_\ell(t)\}$ of the black box to the inputs $\{f_k(t)\}$.
It is in general not possible to fully identify the system $\sigma$ if one does not know any of its elements \cite{ref:Sontag}.
In \cite{ref:QSI}, it is proved for the case where the dynamics in the black box is unitary that the system is identifiable up to unitary equivalence, provided that the system is fully \emph{controllable}.
What if the dynamics is not unitary but governed by the master equation (\ref{master})?
We wilwilll clarify to what extent we can identify the open quantum system and which conditions are required for it.

\textit{Accessibility of open quantum systems.---}
For a fixed choice of the control functions $f_k(t)$, we can formally write the solution to the master equation (\ref{master}) as $
\rho(t)=\mathcal{C}_{t}(\rho_{0})
$, where $\mathcal{C}_{t}$ is a completely positive trace-preserving (CPT) linear map \cite{ref:DynamicalMap-Alicki}. 
Since it takes over the role of the time-evolution operator from closed systems, we call the system \emph{operator controllable} if all time-dependent Markovian CPT maps can be realized by steering $f_k(t)$ \cite{note:DirrHelmke}.
Note that such a set of maps is unequal to the set of all CPT maps \cite{ref:Mixing-Wolf}. Due to the
generally irreversible nature of the dynamics, the operator controllability
is not a very useful concept for open systems 
\cite{note:3}. 
A more useful concept is \emph{accessibility}, which states, roughly
speaking, that all directions can be explored, even if only irreversibly  \cite{note:DirrHelmke}.

Accessibility is rather complicated when talking about directions
of \emph{states}, but has an easy Lie-algebraic characterization when
talking about \emph{operator accessibility}.
One could naively write the time-evolution superoperator $\mathcal{C}_{t}$ and its Lie algebra as complex-valued $d^{2}\times d^{2}$ matrices and define operator
accessibility if a system can explore all $d^{4}$ directions. This
however does not give rise to accessible systems, because directions such
as ``changing the trace'' and ``making Hermitian matrices non-Hermitian''
can never be realized.
To take account of these constraints one usually introduces
an orthogonal basis of Hermitian matrices $(I,\lambda_{1},\ldots,\lambda_{d^2-1})$,
in which density matrices are uniquely expressed as 
\begin{equation}
\rho=\frac{1}{d}I
+\sqrt{\frac{d-1}{2d}}
\sum_{k=1}^{d^2-1}r_{k}\lambda_{k}.
\label{coherence}
\end{equation}
Here, $I$ is the $d\times d$ identity matrix, and the orthogonality
of the traceless Hermitian matrices $\lambda_{j}$ is with respect to the usual
Hilbert-Schmidt product, $\tr\{\lambda_{i}^{\dagger}\lambda_{j}\}=2\delta_{ij}.$
$D=d^{2}-1$ is the effective dimension of the density matrices, and the real $r_{k}$ make up a $D$-dimensional vector $\vec{r}$ called \emph{coherence vector} \cite{note:CoherenceVector}.
In terms of the coherence vector, the time-evolution  
$\rho(t)=\mathcal{C}_{t}(\rho_{0})$
is represented by an affine transformation
\begin{equation}
\vec{r}(t)=V(t)\vec{r}_0+\vec{v}(t),\label{affine}
\end{equation}
where $V(t)$ is a real invertible
\cite{note:4} 
$D\times D$ matrix, while $\vec{v}(t)$ a real vector. 
Again, not every pair $(V(t),\vec{v}(t))$ represents a valid time-evolution since $\mathcal{C}_{t}$ needs to be CPT\@.

A trick is to embed such an affine transformation (\ref{affine}) in a matrix as
\begin{equation}
\left(\begin{array}{c}
\vec{r}(t)\\
1
\end{array}\right)
=\left(\begin{array}{cc}
V(t) & \vec{v}(t)\\
\vec{0}\,^T & 1
\end{array}\right)
\left(\begin{array}{c}
\vec{r}_0\\
1
\end{array}\right).
\end{equation}
Then, the master equation (\ref{master}) is represented by
\begin{equation}
\frac{d}{dt}
\left(\begin{array}{c}
\vec{r}(t)\\
1
\end{array}\right)
=\left(\begin{array}{cc}
A(t) & \vec{b}(t)\\
\vec{0}\,^T & 0
\end{array}\right)
\left(\begin{array}{c}
\vec{r}(t)\\
1
\end{array}\right),
\end{equation}
where 
$
A(t) = A_{0}+\sum_{k}f_{k}(t)A_{k}
$ 
and
$
\vec{b}(t) = \vec{b}_{0}+\sum_{k}f_{k}(t)\vec{b}_{k}
$
correspond to the Lindblad generators in (\ref{master}).
The system Lie algebra is now defined by
\begin{equation}
\mathfrak{l}=\left\langle 
\left(\begin{array}{cc}
A_{0} & \vec{b}_{0}\\
\vec{0}\,^T & 0
\end{array}\right),
\left(\begin{array}{cc}
A_{1} & \vec{b}_{1}\\
\vec{0}\,^T & 0
\end{array}\right),
\ldots,
\left(\begin{array}{cc}
A_{N_{i}} & \vec{b}_{N_{i}}\\
\vec{0}\,^T & 0
\end{array}\right)
\right\rangle _{[\cdot,\cdot]},
\end{equation}
and the system is called operator accessible iff
\begin{equation}
\mathfrak{l}
=
\left(\begin{array}{cc}
\mathbb{R}^{D\times D} & \mathbb{R}^{D}\\
\vec{0}\,^T & 0
\end{array}\right).
\label{eq:access}
\end{equation}

Just
like ``almost all'' systems are controllable in closed dynamics, it
can be shown that the set of accessible open quantum systems
is open and dense. Therefore, accessibility,
opposed to controllability, is a useful premise for considering identifiability (for closed systems these notions coincide).
Below, we will generalize the notion of accessibility to include cases
where, roughly speaking, $\mathfrak{l}$ is smaller, but instead we
have more measurements and state preparations at hand, as would be the case in quantum process tomography.

\textit{Input-output equivalence.---}
In this representation, our outputs $g_\ell(t)$ are expressed as
\begin{equation}
g_\ell(t)
=\tr\{M_{\ell}\rho(t)\} 
=\left(\begin{array}{cc}
\vec{m}_{\ell}^T&
m_\ell^{(0)}
\end{array}\right)
\left(\begin{array}{c}
\vec{r}(t)\\
1
\end{array}\right),
\end{equation}
where $m_\ell^{(0)}=\tr\{M_{\ell}\}/d$ and $\vec{m}_\ell=\sqrt{(d-1)/2d}\tr\{M_\ell\vec{\lambda}\}$. 
The systems are characterized by $\sigma=(\mathcal{L}_{0},\mathcal{L}_{k},\vec{r}_{0},\vec{m}_{\ell},m^{(0)}_\ell)$.
Now, we call two systems equivalent $\sigma\equiv\sigma'$ if they provide
the same outputs whenever the inputs are the same.
This means that when $f_{k}(t)=f'_{k}(t)$ for all $t$ we have $g_\ell(t)=g_\ell'(t)$ for all $t$, i.e.,
\begin{equation}
\left(\begin{array}{cc}
\vec{m}_{\ell}^T&m^{(0)}_\ell
\end{array}\right)
\left(\begin{array}{c}
\vec{r}(t)\\
1
\end{array}\right)
 =
 \left(\begin{array}{cc}
 \vec{m}_{\ell}^{\prime\, T}&
m^{\prime(0)}_\ell
\end{array}\right)
\left(\begin{array}{c}
\vec{r}\,'(t)\\
1
\end{array}\right),
\ \ \forall t.
\label{eqn:InOutEq}
\end{equation}
As in \cite{ref:QSI} there is an algebraic version of this. Slightly stretching
the notation, let us use the same symbol for the generators in (\ref{master})
and their matrix representation,
\begin{equation}
\mathcal{L}_{\bm{\alpha}}
=\left(\begin{array}{cc}
A_{\bm{\alpha}} & \vec{b}_{\bm{\alpha}}\\
\vec{0}\,^T & 0
\end{array}\right).
\label{eqn:LindbladAffine}
\end{equation}
We denote $\mathcal{L}_{\bm{\alpha}}\equiv\mathcal{L}_{\alpha_{K}}\mathcal{L}_{\alpha_{K-1}}\cdots\mathcal{L}_{\alpha_{1}}$,
where $\bm{\alpha}$ is a multi-index of length $K$ with entries
$\alpha_{j}=0,1,\ldots,N_{i}$. Further, we include the case $K=0$
as the identity matrix and denote the corresponding $\bm{\alpha}=\emptyset$.
Then, the condition (\ref{eqn:InOutEq}) algebraically amounts to
\begin{equation}
\left(\begin{array}{cc}
\vec{m}_{\ell}^T&
m^{(0)}_\ell
\end{array}\right)
\mathcal{L}_{\bm{\alpha}}
\left(\begin{array}{c}
\vec{r}_{0}\\
1
\end{array}\right)
=
\left(\begin{array}{cc}
\vec{m}_{\ell}^{\prime T}&
m^{\prime(0)}_\ell
\end{array}\right)
\mathcal{L}_{\bm{\alpha}}'
\left(\begin{array}{c}
\vec{r}_{0}^{\,\prime}\\
1
\end{array}\right).
\label{eq:io}
\end{equation}

\textit{Similarity.---}
We are now ready to extend the arguments in \cite{ref:QSI} for open quantum systems.
Accessibility allows one to expand any vector as
\begin{equation}
\left(\begin{array}{c}
\vec{r}\\
1
\end{array}\right)
=\sum_{\bm{\alpha}}\lambda_{\bm{\alpha}}\mathcal{L}_{\bm{\alpha}}
\left(\begin{array}{c}
\vec{r}_0\\
1
\end{array}\right),
\label{eq:accessibility}
\end{equation}
and we define a linear mapping $T$ by
\begin{equation}
T\left(\begin{array}{c}
\vec{r}\\
1
\end{array}\right)
=\sum_{\bm{\alpha}}\lambda_{\bm{\alpha}}\mathcal{L}'_{\bm{\alpha}}
\left(\begin{array}{c}
\vec{r}_0^{\,\prime}\\
1
\end{array}\right),
\end{equation}
where $\lambda_{\bm{\alpha}}\in\mathbb{R}$ and $\lambda_{\emptyset}=1$.
It is possible to show, under the input-output equivalence (\ref{eq:io}) and the accessibility, that this mapping $T$ is well-defined, even though the decomposition (\ref{eq:accessibility}) is not unique.
We just need to show that for two different decompositions the image of $T$ is the same, and that $T$ is invertible.
We then reach the conclusion that the two systems $\sigma$ and $\sigma'$ which are indistinguishable by the input-output behavior are similar, $\sigma\sim\sigma'$, related by the similarity transformation $T$ \cite{ref:Sontag,ref:QSI}. 
In other words, by looking at the input-output behavior, one can identify the system up to similarity.

In standard quantum process tomography, however, the accessibility is not assumed.
Instead, one tries various initial states and measures various observables to characterize the system.
We can generalize the above scheme in this way, relaxing the accessibility condition.
To this end, assume that the system can be initialized in several  
(unknown but fixed) states $\vec{r}_{j}$. 
Even if the system is not accessible, it is possible to expand any vector as
\begin{equation}
\left(\begin{array}{c}
\vec{r}\\
1
\end{array}\right)
=\sum_{\bm{\alpha},j}\lambda_{\bm{\alpha},j}\mathcal{L}_{\bm{\alpha}}
\left(\begin{array}{c}
\vec{r}_{j}\\
1
\end{array}\right)
\label{eq:rank1}
\end{equation}
with $\lambda_{\bm{\alpha},j}\in\mathbb{R}$ and $\lambda_{\emptyset,j}\ge0$, $\sum_j\lambda_{\emptyset,j}=1$, if a sufficient variety of the initial states $\vec{r}_j$ are available. 
We also need sufficiently many different observables $M_\ell$ that allow one to
write an arbitrary measurement as
\begin{equation}
\left(\begin{array}{cc}
\vec{m}^T&
m^{(0)}
\end{array}\right)
=\sum_{\bm{\alpha},\ell}\mu_{\bm{\alpha},\ell}
\left(\begin{array}{cc}
\vec{m}_{\ell}^T&
m^{(0)}_\ell
\end{array}\right)
\mathcal{L}_{\bm{\alpha}}.
\label{eq:rank2}
\end{equation}
We then define $T$ by
\begin{equation}
T\left(\begin{array}{c}
\vec{r}\\
1
\end{array}\right)
=\sum_{\bm{\alpha},j}\lambda_{\bm{\alpha},j}\mathcal{L}'_{\bm{\alpha}}
\left(\begin{array}{c}
\vec{r}_{j}^{\,\prime}\\
1
\end{array}\right),
\end{equation}
for the vector $\vec{r}$ expanded as (\ref{eq:rank1}), and it is possible to prove the same statement as the one for the accessible case that the two systems $\sigma$ and $\sigma'$ which are indistinguishable by the input-output behavior are related by the similarity transformation $T$.
Summarizing, what we need for identifying a system up to similarity is that both sets
\begin{equation}
\mathcal{L}_{\bm{\alpha}}
\left(\begin{array}{c}
\vec{r}_{j}\\
1
\end{array}\right)
\qquad\text{and}\qquad
\left(\begin{array}{cc}
\vec{m}_{\ell}^T&
m_{\ell}^{(0)}
\end{array}\right)
\mathcal{L}_{\bm{\alpha}}
\label{eq:dualrank}
\end{equation}
have full rank, which is achieved by the accessibility of the system, by a sufficient variety of the initial states $\vec{r}_{j}$ and the measurements $M_\ell$ at hand, or by a scenario between these extremes.

\textit{Structure of $T$.---}
It is easy to see that the similarity transformation $T$ must have the structure
\begin{equation}
T=\left(\begin{array}{cc}
T_{1} & \vec{t}_{2}\\
\vec{0}\,^T & 1
\end{array}\right).\label{eq:structure}
\end{equation}
Since $T$ is invertible, $T_{1}$ must be invertible. Now we come to an essential
difference from the unitary case. At the moment, $T$ transforms
a specific fixed set of Lindbladians into another set of Lindbladians.
In the unitary case, we could use controllability 
 to conclude that in fact \emph{all} unitary
evolutions are transformed into unitary ones, and the similarity transformation $T$ is further constrained to be unitary \cite{ref:QSI}. In the present case, on the other hand, since the
Lindbladian structure is not preserved when generating a Lie algebra,
it is not clear if we can conclude that \emph{all} Lindbladians must
be transformed into Lindbladians. 
We give a negative answer by providing
an explicit example below, in fact showing that the structure of $T$
in (\ref{eq:structure}) cannot be constrained further to unitary.

\textit{Example.---}
Let us consider the following two Lindbladians for the master equation (\ref{master}) for a qubit,
\begin{align}
\mathcal{L}_{0}
&=\left(
\begin{array}{cccc}
-\frac{1}{2} & -1\\
1 & -\frac{1}{2}\\
 &  & -0.9 & -0.89\\
\\
\end{array}
\right),
\\
\mathcal{L}_{1}
&=\left(
\begin{array}{cccc}
-\frac{1}{2} & -1 & -1\\
1 & -\frac{1}{2}\\
1 &  & -0.9 & -0.89\\
\\
\end{array}
\right).
\end{align}
Notice that not all the matrices of the structure (\ref{eqn:LindbladAffine}) are valid Lindbladians: the matrix $\Gamma_{ij}$ (we call it Kossakowski matrix) of a generator for a qubit,
$
\mathcal{L}(\rho)
=-i\sum_{i=1}^3[h_i\sigma_i,\rho]-\frac{1}{2}\sum_{i,j=1}^3\Gamma_{ij}(\sigma_i\sigma_j\rho+\rho\sigma_i\sigma_j-2\sigma_j\rho\sigma_i)
$ [with $\sigma_i$ ($i=1,2,3$) being Pauli matrices], must be positive semi-definite for complete positivity \cite{ref:DynamicalMap-Alicki}.
The eigenvalues of the Kossakowski matrices of the above generators $\mathcal{L}_0$ and $\mathcal{L}_1$ are both given by $\gamma_{0}=\gamma_{1}=\{0.0025,0.0250,0.4475\}$, which are all positive. 
Roughly speaking, the first Lindbladian $\mathcal{L}_0$ describes an amplitude damping process in the presence of a magnetic field in the $z$ direction (but with the elements $-0.9$ and $-0.89$ instead of $-1$) \cite{ref:NielsenChuang}, while the other one $\mathcal{L}_1$ has an additional magnetic field in the $y$ direction. 
This additional field
guarantees that the system is accessible, as is easily confirmed by generating
their Lie algebra numerically (which is $12$-dimensional). Now, consider
the similarity transformation
\begin{equation}
T=\left(\begin{array}{cccc}
1 & 0.01\\
 & 1\\
 &  & 1 & -0.01\\
&&&1\\
\end{array}\right).
\label{eqn:Texample}
\end{equation}
The Lindbladians $\mathcal{L}_0$ and $\mathcal{L}_1$ are transformed by this $T$ to generators whose Kossakowski spectra are given by $\gamma_{0}'=\{0.0047,0.0250,0.4453\}$ and $\gamma_{1}'=\{0.0044,0.0253,0.4453\}$, respectively, which means that they remain valid Lindbladians.
Since the Kossakowski spectra have changed, it is
clear that the two systems cannot be connected by a unitary transformation. 
On the other hand, transforming
a purely unitary component such as
\begin{equation}
\mathcal{L}_{2}=\left(\begin{array}{cccc}
 &  & -1&\\
\\
1\\
\\
\end{array}\right)
\label{eqn:L3}
\end{equation}
is transformed to a generator with a Kossakowski spectrum $\{-0.0035,0,0.0035\}$, which is not physical. 
If we know in advance that we have the control $\mathcal{L}_{2}$, the similarity transformation $T$ in (\ref{eqn:Texample}) is rejected and the similarity transformation connecting the two systems is restricted.
If not, however, we cannot exclude $T$ in (\ref{eqn:Texample}).
To complete the picture, we should also check whether initial states $\vec{r}_{j}$ remain valid by the similarity transformation $T$.
There actually exist such vectors: the completely mixed state is transformed to a valid state,
$\vec{r}_j=(0,0,0)\xrightarrow{\ T\ }(0,0,-0.01)$,
and so are the states around it.
In summary, there actually exists a similarity transformation $T$ connecting two valid, accessible open systems with equal input-output behavior, which are unitarily inequivalent.

\textit{Unitary dynamics without control.---}
Given the rich structure of noisy quantum dynamics, the above fact that indistinguishable open systems are generally not unitarily equivalent is perhaps not surprising. What is remarkable however is that in general this remains the case even under the premise of unitary dynamics.

Assume for simplicity that we apply no control to the system, except for deciding the initial states, run times, and measurements, with the rank condition (\ref{eq:dualrank}) fulfilled by the available states and measurements.
The system is not accessible, and the protocol is similar to standard quantum process tomography, with the main difference that
the reference states and the measurements themselves are not known.
Consider then a Liouvillian $\mathcal{L}_0$ whose spectrum is given by
\begin{equation}
\{0^6,\pm1^2,\pm2^2,\pm3,\pm4,\pm5^2,\pm6^2,\pm7,\pm8,\pm9,\pm10,\pm11\},
\label{eqn:TurnpikeLiouvilleSpectrum}
\end{equation}
where the superscripts denote multiplicities. According to our theorem, each valid black-box model has to have this spectrum, as the models are all related by similarity transformations. 
Let us now assume that we are sure that the dynamics occurring in the black box is unitary, i.e., the true Liouvillian has the structure $\mathcal{L}_0 = -i [H_0,{}\cdot{}\,]$. 
Is it possible 
to identify the Hamiltonian $H_0$ up to unitarity? 
The answer is no.
In fact, two Hamiltonians $H_0$ and $H_0'$ whose spectra are given by
\begin{equation}
\{E_n\}=\{0,1,2,6,8,11\},\ \ %
\{E'_n\}=\{0,1,6,7,9,11\}
\end{equation}
give rise to the same Liouvillian spectrum (\ref{eqn:TurnpikeLiouvilleSpectrum}).
This is an example of the non-uniqueness of the ``turnpike problem'' \cite{TURNPIKE} (interestingly, there are no such systems of dimension smaller than $6$).
We cannot discriminate these two Hamiltonians, which are not unitarily related to each other, from the input-output behavior of the back box.

Yet, having the same spectrum, there must exist a similarity transformation connecting the two Liouvillians $\mathcal{L}_0=-i[H_0,{}\cdot{}\,]$ and $\mathcal{L}_0'=-i[H_0',{}\cdot{}\,]$.
Consider the case where $H_0$ and $H_0'$ are diagonal in the same basis $\{|{n}\rangle\}$.
Then, $\mathcal{L}_0$ and $\mathcal{L}_0'$ are both diagonal on the same operator basis $|{m}\rangle\langle{n}|$ with eigenvalues $-i(E_{m}-E_{n})$ and $-i(E'_{m}-E'_{n})$, respectively.
Because these spectra are equal, they can be related by a permutation of double indices $(m,n)\leftrightarrow(m',n')$.
Furthermore, because the spectrum of Liouvillians is symmetric about zero, $(m,n)\leftrightarrow(m',n')$ implies $(n,m)\leftrightarrow(n',m')$, and in addition the permutation can be chosen such that $(n,n)\leftrightarrow(n,n)$.
Then, the action of this similarity transformation $\mathcal{P}$ on density matrices can be easily seen to be trace-preserving, Hermitianity-preserving,
and unital, $\mathcal{P}(I)=\sum_n\mathcal{P}(|n\rangle\langle n|)=\sum_n|n\rangle\langle n|=I$.
This implies that a ball of states around the maximally mixed state is mapped into another ball of states.
This shows the existence of a valid similarity transformation $\mathcal{P}$ connecting the two systems with the Hamiltonians $H_0$ and $H_0'$.

The non-uniqueness of the turnpike problem means that in the standard framework
of unitary quantum process tomography, without additional knowledge, the input-output behavior only determines the spectrum of Liouvillian, but not of Hamiltonian. The Hamiltonian formalism becomes meaningful only in the presence of further controls used to estimate the system, where the transformation can be represented by a unitary \cite{ref:QSI}.

Finally we remark that the above considerations imply that spectral data cannot be uniquely identified from transition frequencies. This is similar to the non-uniqueness discussed in \cite{ref:SchirmerOi}. However there the ambiguity arrises entirely from not knowing the multiplicity of the transition frequencies.

\textit{Conclusions.---}
We have provided a general framework for the identification of Markovian open quantum systems. The examples disclosed a rich and intriguing structure. In particular we found that systems with different strengths and types of noise can nevertheless display the same input-output behavior, and that even unitary systems with different Hamiltonians can give the same observable data, despite accessibility. Our work sets the frame for further generalizations to non-Markovian systems and systems with feedback dynamics \cite{ref:GutaYamamoto}.

An interesting application of our results would be to identify the input-ouput behavior of the Fenna-Mathews-Olson complex, which is a noisy system in which ultrafast control seems promising \cite{ref:Hoyer}. The question if quantum effects can be observed in this system could be answered unambiguously and without further assumptions if the system is found to be accessible and if each equivalent representation, as characterized by similarity transformations, turns out to be nonclassical.

\textit{Acknowledgements.---}
DB would like to thank David Gross for pointing out reference \cite{TURNPIKE}.
This work is partially supported by the Erasmus Mundus-BEAM Program, and by the Grant for Excellent Graduate Schools from the Ministry of Education, Culture, Sports, Science and Technology (MEXT), Japan.
KY is supported by a Waseda University Grant for Special Research Projects (2013B-147).


\begin{thebibliography}{10}
\bibitem{note:1}
This is not really a constraint: if some superoperator is corresponding to a unitary dynamics, $-i[H,{}\cdot{}\,]$, simply include a further term $i[H,{}\cdot{}\,]$ and an extra control to the summation in (\ref {master}). For non-unitary dynamics negative signs would be unphysical.
\bibitem{note:2}
This can be generalized to POVMs easily, but yields no further insight.
\end{thebibliography}
\end{document}